\documentclass[12pt]{article}
\usepackage{amsmath,amssymb,mathtools,amsfonts,epsfig,graphicx,euscript}
 \usepackage{color}
 \usepackage{etoolbox}
\newcommand{\lGB}{\lambda_{GB}}

\newcommand{\bse}{\begin{subequations}}
\newcommand{\ese}{\end{subequations}}
\newcommand{\be}{\begin{equation}}
\newcommand{\ee}{\end{equation}}
\newcommand{\bea}{\begin{eqnarray}}
\newcommand{\eea}{\end{eqnarray}}
\newcommand{\ba}{\begin{array}}
\newcommand{\ea}{\end{array}}

\begin{document}
\begin{titlepage}
\thispagestyle{empty}

\vspace{2cm}
\begin{center}
\font\titlerm=cmr10 scaled\magstep4
\font\titlei=cmmi10
scaled\magstep4 \font\titleis=cmmi7 scaled\magstep4 {\Large{\textbf{Linearized Holographic Isotropization at Finite Coupling}
\\}}
\setcounter{footnote}{0}
\vspace{1.5cm} \noindent{{
Mahdi Atashi$^{a}$\footnote{e-mail:m.atashi@shahroodut.ac.ir}, Kazem Bitaghsir
Fadafan$^{a}$\footnote{e-mail:bitaghsir@shahroodut.ac.ir
}, Ghadir
Jafari$^{b}$\footnote{e-mail:ghadir.jafari@mail.um.ac.ir
}
}}\\
\vspace{0.8cm}

{\it $^{a}$Physics Department, Shahrood University of Technology,\\ P.O.Box 3619995161 Shahrood, Iran\\}
{\it $^{b}$School of Physics,Institute for Research in Fundamental Sciences (IPM),\\P.O. Box 19395-5531, Tehran, Iran}

\vspace*{.4cm}

\end{center}
\vskip 2em
\setcounter{footnote}{0}
\begin{abstract}
We study holographic isotropization of an anisotropic homogeneous non-Abelian strongly coupled plasma in the presence of Gauss-Bonnet corrections. It was verified before that one can linearize Einstein's equations around the final black hole background and simplify the complicated setup. Using this approach, we study the expectation value of the boundary stress tensor. Although we consider small values of the Gauss-Bonnet coupling constant, it is found that finite coupling leads to significant increasing of the thermalization time. By including higher order corrections in linearization, we extend the results to study the effect of the Gauss-Bonnet coupling on the entropy production on the event horizon.

\end{abstract}

\end{titlepage}

\section{Introduction}
Regarding the experiments at Relativistic Heavy Ion Collisions (RHIC) and LHC, a strongly$-$coupled quark$-$gluon plasma (QGP) has been produced by collision of heavy ions (see review \cite{CasalderreySolana:2011us}). The fast thermalization and the applicability of viscous hydrodynamics about $1$ fm/c or less after the collision of ions is puzzlingly small. There are no known
quantitative methods to study such strongly$-$coupled process from perturbation theory even by lattice
simulations. This could be good motivation to study thermalization process in strongly$-$coupled medium from holographic approach. Using the holographic techniques \cite{CasalderreySolana:2011us,Maldacena:1997re,Gubser:1998bc,Witten:1998qj,
Witten:1998zw} has yielded many important insights into
the dynamics of strongly coupled non-Abelian theories. In this approach gravity in $AdS_5$ space is related to the conformal
field theory on the four-dimensional boundary \cite{Witten:1998qj}.
It was also shown that an $AdS$ space time with a black hole is dual to a
conformal field theory (CFT) at finite temperature \cite{Witten:1998zw}.\\

The thermalization from the gravity side means the process where a bulk background achieves the formation of a static black hole \cite{Chesler:2008hg,Chesler:2010bi,Heller:2011ju,Heller:2012je}. One may call this stage as hydrodynamization where the system approaches a new phase which the dynamics of system is given by the hydrodynamic equations. The details of this phenomena can be understood from fluid-gravity duality \cite{Bhattacharyya:2008jc}. However, the process needs solving Einstein's equation numerically. Fortunately doing the numerics in the AdS space time is easier \cite{Chesler:2013lia}.\\

In this paper, we study the holographic isotropization of a homogeneous non-Abelian strongly coupled in the presence of Gauss-Bonnet corrections. As a general result of the
holography, the effects of finite but large
't Hoof coupling $\lambda$ in the boundary gauge field theory are captured by
adding higher derivative terms in the corresponding geometry. \footnote{The 't Hoof coupling
$\lambda$ is related to the curvature radius of the $AdS_5$ space time and
$S_5$ sphere $(L)$, and the tension of the string $(\frac{1}{2\pi \alpha'})$ by this relation $\sqrt{\lambda}=\frac{L^2}{\alpha'}$.}

The curvature squared terms like the Gauss-Bonnet corrections are common from the sense that comes from string theory and also that in the resulting action there is no ghost. The effect of these corrections on the different aspects of heavy quarks in the QGP has been studied in \cite{Fadafan:2013coa,Fadafan:2015kma,Fadafan:2012qy,Fadafan:2011gm,AliAkbari:2009pf,Fadafan:2008uv,Fadafan:2008gb}. See also related studies in this subject in \cite{Zhang:2015hkz,Misobuchi:2015ioa,Ficnar:2013qxa,Zhang:2012jd,Arnold:2012uc}.\\

An understanding of how the isotropization process of a non-Abelian strongly coupled plasma is affected by considering finite coupling corrections may be essential for theoretical predictions \cite{Grozdanov:2016vgg}. It may be crucial to understand if the fast thermalization depends on these corrections. It would be important to notice that most of the such analysis have been done for gauge theories with an Einstein gravity dual in the limit of $\lambda \rightarrow \infty$ \cite{Chesler:2008hg,Heller:2012km}. Then it would be natural to ask if the main results of such analysis can be changed at finite $\lambda$. One important observation in this case is violation of the bound on the shear viscosity to entropy density, $\frac{\eta}{s}$ in CFTs dual to Gauss-Bonnet gravity \cite{Brigante:2007nu}. Although, the theory may be inconsistent regarding microcausality \cite{Camanho:2009vw}. \\

Study of short isotropization time is an example of far from equilibrium phenomena which requires numerical solution of gravity dynamics with non trivial initial conditions. In this case one needs to solve the full non$-$linear Einstein's equations, numerically. As it was pointed out doing such initial problem in asymptotically AdS geometries is much tractable and sometimes is referred as numerical holography, see \cite{vanderSchee:2014qwa}. In addition of finite difference methods, pesudospectral methods are also used for solving non linear Einstein's equation in numerical holography \cite{Grandclement:2007sb}.  One may find solved examples as turbulence in 2D fluids \cite{vanderSchee:2013pia}, collisions of shock waves \cite{Chesler:2010bi,Casalderrey-Solana:2013aba,Casalderrey-Solana:2013sxa} or boost invariant expansion \cite{Chesler:2009cy,Heller:2012je} and Wilson loop evolution \cite{Ali-Akbari:2015ooa}.\\

Following the approach of \cite{Heller:2012km,Heller:2013oxa}, we consider isotropization of a homogeneous non-Abelian plasma in a four-dimensional CFT in flat Minkowski space time. In this case, one should introduce far from equilibrium states by considering a large number of initial states in the absence of external sources which does not need to deform the boundary gauge theory.\footnote{One finds different approach in \cite{Chesler:2008hg} by turning on an anisotropic source which pumps energy and momentum into the CFT vacuum to create a far from equilibrium state.} Recently,  this approach has been followed in \cite{Ali-Akbari:2016sms} by studying Einstein's  general  relativity  coupled  to  a  massive  scalar  matter   field. In this case, far from equilibrium initial  states are  described  by  a  non-trivial scalar matter field plus an anisotropic metric ansatz in dual gravity theory. In the gravity side, the time evolution of each state is given by numerically solving Einstein's equation.\\

As \cite{Heller:2012km,Heller:2013oxa}, we consider the amplitude expansion by linearizing the Einstein's equations on top of the black hole background. This is the only existing approximation scheme apart from the studying of thermalization in the AdS$-$Vadia black hole background. It was shown that such approximation describes very well the one$-$point function of the boundary energy-momentum tensor. Here, we extend this observation to the case of finite coupling corrections. The Gauss-Bonnet correction term which is quadratic in the curvature yields second order equations of motion and possesses an exact black brane solution with AdS asymptotics. One should notice that such solutions might not be obtained from string theory side.\\

Recently, the structure of thermal energy-momentum tensor correlators at finite coupling has been studied in \cite{Grozdanov:2016vgg}. They investigate corrections to the eigenvalues of the linearized Einstein's equations, i.e. the quasinormal spectra of black holes in the presence of higher derivative Gauss$-$Bonnet gravity and $R^4$ terms. It is wellknown that the least damped non-hydrodynamic modes play an important role in the study of relaxation phenomena. Also using numerical holography reveal that the hydrodynamic stage, i.e hydrodynamization is reached before isotropization. The related time scales are determined by the lowest quasinormal frequency \cite{Buchel:2015saa,Janik:2015waa,Janik:2015iry,Attems:2016ugt,Janik:2016btb}. One finds in \cite{Grozdanov:2016vgg} that if the known relation between transport coefficients and the relaxation times from kinetic theory exists at Gauss$-$Bonnet theory. It is found that the ratio of a transport coefficient to the relaxation time shows an extrapolation from strong coupling to the
kinetic theory results at weak coupling. Also, it is shown that the quasinormal spectrum depends on the behavior of $\eta/s$ at finite coupling.\\

Our purpose in this paper is to explore the applicability of linearized gravity equations further, especially for the case of higher derivative corrections where solving the partial differential equations are complicated. We start
by studying the black hole background in the Gauss$-$Bonnet gravity and produce the non$-$linear equations.\\

One can probe the gravitational dynamics of the isotropization process, by different field theory observables. Here, we first focus on the time evolution of the one$-$point function of the stress tensor and next we study the entropy production. Although for far from equilibrium states defining the entropy density does not precise definition, we define it as the change of the area density of the event horizon. Other examples to probe the system is study of expectation value of local operators, entanglement entropy and Wilson loops.\\

\textbf{Note added:} While this paper was in the final stages of preparation, the related papers \cite{Andrade:2016rln} and \cite{Grozdanov:2016zjj} appeared on arXiv. The work of \cite{Andrade:2016rln} studies a similar idea about holographic isotropization of homogeneous, strongly coupled, non-Abelian plasmas in Gauss-Bonnet gravity with a negative cosmological constant. They numerically solve the linearized equations by the quasinormal mode expansion which is different from our approach. Interestingly, it is shown that Gauss-Bonnet corrections increase the isotropization time of strongly coupled plasma. Also, the time evolution of the pressure anisotropy with the Gauss-Bonnet correction is shifted. Our results are in perfect agreement with \cite{Andrade:2016rln}. In \cite{Grozdanov:2016zjj}, the effect of the Gauss-Bonnet coupling on the non equilibrium dynamics of the debris of two shock-wave collisions has been studied, holographically.\\

This paper is organized as follows. In sections two, we review the linearized isotropization approach by holography. We also explore non$-$linear Einstein's equations in this section. We study the holographic setup in the presence of Gauss-Bonnet coupling corrections in section three. In this section we derive the nested form of the non linear Einstein$-$Gauss$-$Bonnet's equations. The linearizing Einstein$-$Gauss
$-$Bonnet's equations have been done in section four. By including higher order corrections, we extend the results to study the entropy production on the event horizon in section five. In the last section we summarize our results.

\section{Review of Linearized Isotropization} 
In this section we review the approach of \cite{Heller:2012km,Heller:2013oxa}. We study an anisotropic and homogeneous state of strongly coupled $\mathcal{N}=4$ SYM plasma in four dimensions. Consider an initial state with a time dependent pressure anisotropy which leads to a non-equilibrium state. Because of no other time dependent source in the field theory, the boundary metric is flat. For simplicity, the rotational symmetry imposed in two of the spapcelike directions.

The most general form of the dual background metric ansatz is given in the ingoing Eddington-Finkelstein coordinates as follows
\begin{equation}
\label{mansatz}
ds^2 = 2 dt dr - A\left(r,t\right) dt^{2} + \Sigma\left(r,t\right)^{2} e^{-2 B\left(r,t\right)} dx_{L}^{2} + \Sigma\left(r,t\right)^{2}e^{B\left(r,t\right)} d\mathbf{x}_{T}^{2} \,,
\end{equation}
where $r$ is the radial coordinate and the boundary space coordinates are $(x_L,\vec{x}_T)$ with the rotational symmetry in the transverse directions $\vec{x}_T$. The boundary is also located at infinity. The unknown functions $A, \Sigma$ and $B$ are determined by numerical holography as we will discussed later. In these coordinates, null hypersurfaces are given by constant time slices similar to radial ingoing null geodesics.

The metric ansatz \eqref{mansatz} should solve the non$-$linear Einstein's equation
\be
R_{ab}-1/2\,R\,g_{ab}-6/L^2 g_{ab}=0.
\ee
Here, the radius of the AdS space time is given by $L$. We work in units of
$L = 1$ henceforth. Replacing the ansatz in Einstein's equations, one finds the near boundary expansion of the metric functions  as
\begin{subequations}
\label{nearbdryexpansions}
\begin{eqnarray}
\label{a}
A\left(r,t\right)&=& r^2 + \frac{a_{4} }{r^{2}} - \frac{2b_{4}(t)^{2} }{7 r^{6}}  + \cdots \,, \\ [2mm]
\label{b}
B\left(r,t\right)&=&\frac{b_{4}(t)}{r^4} + \frac{\partial_{t} b_{4}(t)}{r^5}  +  \cdots  \,, \\ [2mm]
\label{sigma}
\Sigma\left(r,t\right) &=& r - \frac{b_{4}(t)^{2}}{7 r^{7}}  + \cdots \,, \,\,\,\,\,\,\,
\end{eqnarray}
\end{subequations}
The metric ansatz \eqref{mansatz} enjoys the residual gauge freedom from $r \rightarrow r+f(t)$. This freedom is fixed by considering the near boundary expansion of $A(r,t)$ in \eqref{a} so that the term proportional to $r$ vanishes.

The unknown near boundary coefficients of $a_4$ and $b_4(t)$ should be determined from solving the time dependent background differential equations with suitable initial conditions. From the AdS/CFT correspondence, they are wellknown as normalizable modes and using the holographic renormalization method are identified with the stress tensor of the boundary gauge theory.

The traceless and conserved stress tensor of the boundary theory is given by
\be
 T_{ab} \propto diag\,[ \mathcal{E},\mathcal{P}_L(t),\mathcal{P}_T(t),\mathcal{P}_T(t)] ,
\ee
where $ \mathcal{E}$ is proportional to the energy density which does not change in this setup and would be as an initial condition of the non-equilibrium system. The longitudinal and transverse pressures are given by $\mathcal{P}_L(t)$ and $\mathcal{P}_T(t)$, respectively. The time dependent anisotropy is introduced by $\Delta\,\mathcal{P}(t)$ as follows
\be
\mathcal{P}_L(t)=\frac{\mathcal{E}}{3}-\frac{2\,\Delta\,\mathcal{P}(t)}{3},\,\,\,\,\,\,\,\,\,\,\mathcal{P}_T(t)=\frac{\mathcal{E}}{3}+\frac{2\,\Delta\,\mathcal{P}(t)}{3}.
\ee
For the case of $SU(N_c)$ $\,\mathcal{N}=4$ SYM, the relations between coefficients of $a_4$ and $b_4(t)$ are
\be
\mathcal{E}=\frac{-3a_4}{4},\,\,\,\,\,\,\Delta\,\mathcal{P}(t)=3\,b_4(t).
\ee

When the system reaches to the equilibrium, one may define the temperature of the system as $T$. Also, the energy density is given in terms of $T$ by $\mathcal{E}=\frac{3\,\pi^4\,T^4}{4}$. In this situation the metric describes the AdS-Schwarzschild black brane solution where the metric functions takes the following form
\be
A(r,t)=r^2\left(1-\frac{\pi^4\,T^4}{r^4}\right),\,\,\,\,\,\Sigma(r,t)=r,\,\,\,\,\,B(r,t)=0.
\ee
The isotropization time $t_{iso}$, defines as the time after which the $\Delta\,\mathcal{P}(t)$ remains small with respect to $\mathcal{E}$. Approximately, we adopt the following inequality
\be\label{tiso}
\frac{\Delta\,\mathcal{P}(t>t_{iso})}{\mathcal{E}}\, \leq\,0.1.
\ee

\subsection{Non$-$linear Einstein's equations} 
To have the Einstein's equations for the metric background \eqref{mansatz}, one should define derivatives along the ingoing radial null geodesics and temporal derivatives as
\be
h'\equiv \partial_r h,\,\,\,\,\,\,\,\dot{h}\equiv \partial_t h+\frac{1}{2}A \partial_r h,
\ee
Therefore, the Einstein's equations take the following nested form
\begin{subequations}\label{ODE}
\begin{eqnarray}
\label{Seq}
0 &=& \Sigma \, (\dot \Sigma)' + 2 \Sigma' \, \dot \Sigma - 2 \Sigma^2\,,
\\ \label{Beq}
0 &=& \Sigma \, (\dot B)' + {\textstyle \frac{3}{2}}
    \big ( \Sigma' \dot B + B' \, \dot \Sigma \big )\,,
\\  \label{Aeq}
0 &=& A'' + 3 B' \dot B - 12 \Sigma' \, \dot \Sigma/\Sigma^2 + 4\,,
\\  \label{Cr}
0 &= & \ddot \Sigma
    + {\textstyle \frac{1}{2}} \big( \dot B^2 \, \Sigma - A' \, \dot \Sigma \big)\,,
\\ \label{Ct}
0 &=& \Sigma'' + {\textstyle \frac{1}{2}} B'^2 \, \Sigma\,,
\end{eqnarray}
\end{subequations}
Now one should consider initial time slice of the geometry and study numerically the bulk space time to find the dual stress tensor. Two last equations in \eqref{ODE} are constrains on the initial states. There is a nested algorithm for solving  \eqref{ODE} in which one should use evolution equations \eqref{Seq}, \eqref{Beq} and \eqref{Aeq} at each time step. There are some conditions on the initial states to obtain a far from equilibrium state. Also one should check that singularities must be hidden inside the event horizon. A procedure introduces for choosing $B(r)$ and $\mathcal{E}$ to produce a class of far from equilibrium states. We will derive \eqref{ODE} in the presence of Gauss-Bonnet corrections in the next section. \\

Changing the variable from $r$ to $z=1/r$ is more favorite in the numerical holography. In this case the boundary is located at $z=0$ and the black brane creates at $z=1$. To have a very moderate grid in the $z$ direction, using spectral method is better. The spectral methods in the context of numerical general relativity has been reviewed in \cite{Wu:2011yd}. \footnote{The numerical code is written in \emph{Mathematica} and one can find the source files with specific examples in https://sites.google.com/site/wilkevanderschee/ads-numerics.}\\

Based on the outcome of the numerical simulations one finds that by studying the gauge theory quantity $\frac{\Delta \mathcal{P}(t)}{\mathcal{E}}$ for different initial profiles of $B(r,t)$, the behavior of $t_{iso}$ from  \eqref{tiso}, quantitative. One finds the fast thermalization, i.e the $\Delta \mathcal{P}(t)$ quickly relaxes to zero. The longest isotropization times can be archived by considering the profiles for $B(r,t)$ which is localized close to the horizon. In this case the out going wave packet propagate from the horizon to the boundary and finally fall into the black hole. The range of maximum value of $t_{iso}$ are about $\frac{1.1}{T}\,-\,\frac{1.2}{T}$. We will check that how the longest thermalization time changes at finite coupling.

\subsection{Linear  Einstein's equations} 
Holographic isotropization can be simplified by linearizing Einstein's equations around the final black brane solution, i.e the AdS-Schwarzschild black brane in this case. The linearizing Einstein's equations is interpreted as an amplitude expansion on top of the AdS-Schwarzschild black brane.

By considering the parameter of the expansion as $\alpha$, one expands the metric functions as
\begin{subequations}
\label{linearizedmetricfuncs}
\begin{eqnarray}
\label{linearize.a}
A\left(t,z\right)&=& \frac{1-z^4}{z^2} +\alpha\, \delta A^{(1)}(t,z)+\mathcal{O}\left(\alpha ^2 \right) \,, \\ [2mm]
\label{linearize.b}
B\left(t,z\right)&=&\alpha\, \delta B^{(1)}(t,z)+\mathcal{O}\left(\alpha ^2 \right)  \,, \\ [2mm]
\label{linearize.c}
\Sigma\left(t,z\right) &=& \frac{1}{z}+\alpha\, \delta \Sigma ^{(1)}(t,z)+\mathcal{O}\left(\alpha ^2 \right) \,. \,\,\,\,\,\,\,
\end{eqnarray}
\end{subequations}
Regarding the close-limit approximation in \cite{Anninos:1995vf,Price:1994pm}, the initial far from equilibrium states will not be small perturbations of the AdS-Schwarzschild black brane. Inserting these perturbations into Einstein's equations, one finds that $\delta A^{(1)} \left(t,z\right)=0$ and $\delta \Sigma ^{(1)} \left(t,z\right)=0$. Also evolution equation for $\delta B ^{(1)}\left(t,z\right)$  is given by the following first order time partial differential equation
\begin{align}\label{Bequation}
&\left(z^4+3\right) \partial_z\delta B+z \left(z^4-1\right) \partial^2_z\delta B-3 \partial_t\delta B+2 z \partial_t\partial_z\delta B=0.
\end{align}
The initial condition to solve this equation is
\begin{equation}\label{iniforB}
\delta B^{(1)}\left(t=0,z\right)=B\left(t=0,z\right).
\end{equation}
The energy density $\mathcal{E}$ is also constant in this setup, which is equal to $3/4$. For stability computations, the metric function $B(t,z)$ is regularized as
\be \label{regforB}
\delta B_{reg}^{(1)}\left(t,z\right)=\frac{1}{z^3}\delta B^{(1)}\left(t,z\right)
\ee
which satisfies the condition $\delta B_{reg}^{(1)}\left(t,z=0\right)=0$. The other boundary condition is given inside the event horizon of AdS-Schwarzschild black brane.\\

Solving \eqref{Bequation}, is the main part of analysis of linearized holographic isotropization. By finding its solution, one can study the pressure anisotropy $\Delta \mathcal{P}(t)$. Then the quantity $\frac{\Delta \mathcal{P}(t)}{\mathcal{E}}$ can be found as the leading order dynamics of the process.

By studying 800 far from equilibrium initial state and solving \eqref{Bequation}, it is found that the linearized approach predicts $t_{iso}$ with a $20$ percent accuracy. It is a natural question if including higher order expansion terms leads to closer results. In this case one should consider $\delta\,B^{(3)}, \delta\,\Sigma^{(3)}$ and $\delta\,A^{(4)}$.

By comparing the results of linear and non linear analysis one finds a surprising result that leading order equation \eqref{Bequation} did not result a large effect on the stress tensor of the boundary theory. Briefly, the careful comparisons of linear and non linear approaches show that
\begin{itemize}
\item  At early times the pressure anisotropy have the same behavior. That is because of the fact that the near boundary dynamics is approximately linear.
\item  The pressure anisotropy only differs at transient time because in this case the signal propagates from the interior of the bulk geometry.
\end{itemize}

Therefore one concludes that linear analysis yields a very good approximation framework for studying holographic isotropization.  Also it leads to a very significant simplification. It is desirable to apply this framework in the presence of a complicated setup like considering the Gauss-Bonnet corrections.
\section{Holographic setup at finite coupling} 
In this section we consider finite coupling corrections on the
thermalization process. As it was explained in the introduction section, an understanding of how the dynamics changes by
these corrections may be essential for theoretical predictions.\\

In five dimensions, we consider the theory of gravity with quadratic
powers of curvature $R^2$ as Gauss-Bonnet theory. In this case the derivatives in the equations of motion are second order. The Gauss-Bonnet theory is an example of more general Lovelock theories where the usual difficulties of considering higher derivative terms like instability is absent. Hence, they are interesting for studying non perturbative effects in the presence of higher derivative corrections. An important example of such study is violation of $\eta/s$ bound in the Gauss-Bonnet gravity as \cite{Brigante:2007nu}
\be
\frac{\eta}{s}=\frac{1}{4\pi}\left(1-4\,\lambda_{GB}\right).
\ee
where $\lambda_{GB}$ is the dimensionless parameter and is related to the scale of the higher derivative correction $L_{GB}$ and L as $\lambda_{GB}=\frac{L_{GB}^2}{L^2}$.\\

The action we consider for the bulk theory takes the following form
\begin{equation}
S=\int dx^5 \sqrt{-g}\left(R+\lambda_{GB}\,\mathcal{L}_{GB}\right),
\end{equation}
where\\
\begin{equation}
\mathcal{L}_{GB}=
R_{cdef}R^{cdef}-4R_{ab}R^{ab}+R^2.
\end{equation}
The exact AdS black hole solutions
and their thermodynamic properties in Gauss-Bonnet gravity were
discussed in \cite{Cai:2001dz,Nojiri:2001aj,Nojiri:2002qn}.

The AdS black hole solution is given by%
\begin{equation}
ds^2=-N \,u^2\, h(u)\,dt^2+\frac{1}{u^2
h(u)}\,dr^2+u^2\,d\vec{x}^2\label{GBmetric},
\end{equation}
where
\begin{equation}
h(u)= \frac{1}{2\lambda_{GB}}\left( 1-\sqrt{1-4 \lambda_{GB}\left(
1-\frac{u_h^4}{u^4} \right)}\,\right),
\end{equation}
and the Hawking temperature is given by
\begin{equation}\label{TGB}
 T_{}=\sqrt{N}\,\frac{u_h}{\pi L^2}.
\end{equation}

In (\ref{GBmetric}), $N= \frac{1}{2}\left(
 1+\sqrt{1-4 \lambda_{GB}} \right)$ is an arbitrary constant and specifies
the speed of light of the boundary field theory. It has been chosen to
be unity. Beyond $\lGB<1/4$ there is no vacuum AdS solution and
one cannot have a CFT. However by studying the relation between positivity of the energy constraints in CFT's and causality in their gravity dual description, one finds the constraints imposed on the higher curvature terms \cite{Brigante:2008gz,deBoer:2009pn,Camanho:2009hu,Camanho:2009vw,Buchel:2009sk}.
Then the constraints lead to the bound on the Gauss-Bonnet coupling as
\be
-7/36<\lGB<9/100.
\ee
\\

The metric \eqref{mansatz} has to solve the Einstein's equations with the negative cosmological constant and Gauss-Bonnet higher derivative terms as

\begin{eqnarray}\label{eom}
 R_{ab} +4g_{ab}+ \lGB \,\mathcal{H}_{ab}=0\,,
\end{eqnarray}

here, $\mathcal{H}_{ab}$ is given by

\begin{align}\label{H}
\mathcal{H}_{ab}=-4 R_{ac}
R_{b}{}^{c} + 2 R_{ab} R - 4 \
R_{cd}
R_{a}{}^{c}{}_{b}{}^{d} + 2 \
R_{a cde} R_{b}{}^{cde}.
\end{align}

Having Gauss-Bonnet corrections and the asymptotic AdS space with effective radius $ L_c $, one finds again the near boundary expansion of metric components as

\begin{subequations}
     	\label{nearbdryexpansionsgb}
     	\begin{eqnarray}
     	\label{nearbdryexpansionsgb.a}
     	A\left(r,t\right)&=& \frac{r^2}{L_c^2} + \frac{a_{4} }{r^{2}} +\frac{2b_{4}(t)^{2} \left(6 L_c^2-7\right) -7 a_4^2 L_c^4
     		\left(L_c^2-1\right)}{7r^6 L_c^2 \left(2
     		L_c^2-1\right)}  + \cdots \,, \\ [2mm]
     	\label{nearbdryexpansionsgb.b}
     	B\left(r,t\right)&=&\frac{b_{4}(t)}{r^4} +L_c^2 \frac{\partial_{t} b_{4}(t)}{r^5}  +  \cdots  \,, \\ [2mm]
     	\label{nearbdryexpansionsgb.c}
     	\Sigma\left(r,t\right) &=& r +(\frac{6 L_c^2-7}{2
     		L_c^2-1}) \frac{b_{4}(t)^{2}}{7 r^{7}}  + \cdots \,, \,\,\,\,\,\,\,
     	\end{eqnarray}
     \end{subequations}
which at $ \lambda_{GB}\to 0 $ or $ L_c\to 1 $ reduce to relations \eqref{nearbdryexpansions}. Using this boundary expansion, one can get the following expression for expectation value of stress tensor in  dual theory \footnote{The boundary stress tensor is presented in \cite{Brihaye:2008xu}. One may also use the results of \cite{Jahnke:2014vwa}.}

   \begin{equation}\label{24}
   \hat{T}_{ab}\propto \text{diag}\big(-\frac{3}{2}a_4,\frac{b_4(t)}{L_c^2}-\frac{1}{2}a_4,\frac{b_4(t)}{L_c^2}-\frac{1}{2}a_4,-2\frac{b_4(t)}{L_c^2}-\frac{1}{2}a_4\big).
   \end{equation}
So again the pressure anisotropy is obtained from asymptotic behavior of $ B(r,t) $:
\begin{equation}
\delta \mathcal{P} (t) \propto 3 b_{4}(t)/L_c^2.
\end{equation}
where the coefficient of proportion depends on the effective AdS radius or the $ \lambda_{GB} $. A detailed discussion
of how the initial states depend on the $ \lambda_{GB} $ will be presented in the next section.

The evolution of $ b_4(t) $ and so pressure anisotropy can not be obtained from near boundary expansion and we must solve non-linear bulk equations.

\subsection{Non$-$linear Einstein$-$Gauss$-$Bonnet's equations}
In this subsection for the first time we derive the non$-$linear Einstein's equations in the Gauss$-$Bonnet gravity. We obtain the equation of motion for the metric ansatz \eqref{mansatz} in terms of derivatives along the ingoing and outgoing radial null geodesics. Finally, the Einstein$-$Gauss$-$Bonnet's equations can be presented as the following nested form

\begin{itemize}
\item  The equation \eqref{Seq} changes as
\begin{align}\label{GBa1}
&\left(-2\Sigma ^2+\Sigma  \dot{\Sigma '}+2 \dot{\Sigma }
\Sigma '\right)+\frac{\textcolor{blue}{\lambda _{\text{GB}}}}{\Sigma} \bigg\{\dot{B}^2 \Sigma  \Sigma '
\left(\Sigma  B'+2 \Sigma '\right)+2 \dot{\Sigma }
\left(\Sigma  B' \left(\Sigma  \dot{B'}+\dot{\Sigma }
B'\right)-4 \dot{\Sigma '} \Sigma '\right)\nonumber\\&+\dot{B}
\Sigma ^2 \left(B' \left(\dot{\Sigma } B'+2
\dot{\Sigma '}\right)+2 \dot{B'} \left(\Sigma
B'+\Sigma '\right)\right)\bigg\}  =0.
\end{align}

\item  In the presence of $\lGB$, the equation \eqref{Beq} becomes
\begin{eqnarray}\label{GBb1}
&& \Sigma  (2
\Sigma  \dot{B'}+3 \dot{\Sigma } B'+3 \dot{B} \Sigma
')-\textcolor{blue}{ \lambda _{\text{GB}}} \bigg\{2 \dot{\Sigma }
\Sigma  A'' B'+2 \dot{\Sigma } \Sigma  A' B''+4
\dot{\Sigma } A' B' \Sigma '-\dot{\Sigma } \Sigma  A'
B'^2\nonumber\\&&+2 \dot{B} (\Sigma  (A''
(\Sigma  B'+\Sigma ')+A' (\Sigma
B''+2 B' \Sigma '+\Sigma ''))+\dot{\Sigma
	'} (4 \Sigma '-2 \Sigma  B')\nonumber\\&&-4 \dot{\Sigma
} (\Sigma  B''+B' \Sigma '+\Sigma
''))-4 \ddot{B} \Sigma ^2 B''-4
\ddot{\Sigma } \Sigma  B''+4 \Sigma ^2
\dot{B'}^2-8 \ddot{B} \Sigma  B' \Sigma '\nonumber\\&&+8
\Sigma  \dot{B'} (\dot{\Sigma } B'+\dot{B}
\Sigma '+\dot{\Sigma '})+8 \dot{\Sigma } B'
\dot{\Sigma '}-8 \ddot{\Sigma } B' \Sigma '+2
\ddot{\Sigma } \Sigma  B'^2+4
\dot{\Sigma }^2 B'^2\nonumber\\&&-4 \ddot{B}
\Sigma  \Sigma ''+2 \dot{B}^2 (\Sigma  \Sigma
''+2 \Sigma
'^2)\bigg\}=0.
\end{eqnarray}

\item   In the presence of $\lGB$, the equation \eqref{Aeq} becomes
\begin{align}\label{em5}
&3
\dot{B} B'+\frac{8 \dot{\Sigma '}}{\Sigma }+\frac{4
	\dot{\Sigma } \Sigma '}{\Sigma ^2}-12+A'' \left(1-\textcolor{blue}{\lambda _{\text{GB}}} (\frac{8 \dot{\Sigma
	} \Sigma '}{\Sigma ^2}-2 \dot{B}
B')\right)\nonumber\\&-\frac{\textcolor{blue}{\lambda _{\text{GB}}}}{\Sigma ^2} \bigg\{4
\dot{\Sigma } \Sigma  A' B'^2+\dot{B}
\left(B' \left(\Sigma  A' \left(\Sigma  B'-4 \Sigma
'\right)+16 \Sigma  \dot{\Sigma '}+8 \dot{\Sigma }
\Sigma '\right)-2 \Sigma  B'' \left(\Sigma  A'-4
\dot{\Sigma }\right)\right)\nonumber\\&+8 \dot{\Sigma } A' \Sigma
''+4 \ddot{B} \Sigma ^2 B''-2 \dot{B}^2
\left(\Sigma  \left(\Sigma  B''+4 \Sigma ''\right)+2
\left(\Sigma '\right)^2\right)-4 \Sigma ^2
\dot{B'}^2\nonumber\\&-2 \ddot{B} \Sigma ^2
\left(B'\right)^2+4 \Sigma  \dot{B'} \left(\dot{B}
\left(\Sigma  B'-2 \Sigma '\right)-2 \dot{\Sigma }
B'\right)+8 \ddot{B} \Sigma  B' \Sigma '-8
\ddot{\Sigma } \Sigma  \left(B'\right)^2\nonumber\\&-4
\dot{\Sigma }^2 \left(B'\right)^2-16 \ddot{\Sigma
} \Sigma ''+16 \dot{\Sigma '}^2\bigg\}=0.
\end{align}

\item  The equation \eqref{Cr} changes to
\begin{align}\label{GBd1}
&\frac{1}{2} \left(\dot{B}^2 \Sigma -\dot{\Sigma }
A'\right)+\ddot{\Sigma } \left(1-\textcolor{blue}{\lambda
_{\text{GB}}} (\frac{8 \dot{\Sigma } \Sigma
	'}{\Sigma ^2}-2 \dot{B} B')\right)\nonumber\\&-\frac{\textcolor{blue}{\lambda _{\text{GB}}}}{\Sigma
	^2} \bigg\{-2 \dot{B} \Sigma
\left(-\Sigma  \dot{\Sigma } A' B'+\ddot{B}
\Sigma  \left(\Sigma  B'+\Sigma '\right)+2
\dot{\Sigma }^2 B'\right)\nonumber\\&+\dot{B}^2 \Sigma ^2
\left(A' \left(\Sigma  B'+\Sigma '\right)-3
\dot{\Sigma } B'\right)-2 \dot{\Sigma } \left(2
\dot{\Sigma } A' \Sigma '+\ddot{B} \Sigma ^2
B'\right)+\dot{B}^3 \Sigma ^2 \Sigma '\bigg\}=0.
\end{align}

\item The equation \eqref{Ct} becomes
\begin{eqnarray}\label{em1}
&&\Sigma^{''}\left(2+4 \textcolor{blue}{\lambda _{\text{GB}}} (\dot{B} B'-4
\frac{\dot{\Sigma } \Sigma '}{\Sigma^2}) \right)=-
\Sigma B'^2+2   \frac{\textcolor{blue}{\lambda
	_{\text{GB}}}}{\Sigma} \bigg(\dot{B} -B' \Sigma ' \left(3
\Sigma  B'+4 \Sigma '\right)\nonumber\\&&-2 \Sigma  B''
\left(\Sigma  B'+\Sigma '\right)+\Sigma
\dot{\Sigma } B' \left(B'^2-2
B''\right)\bigg).
\end{eqnarray}

\end{itemize}
As it is clear the Einstein-Gauss-Bonnet's equations are so complicated set up, and one can not use the prescription of \cite{Heller:2012km,Heller:2013oxa}. It would be interesting to use numeric techniques for solving these fully non$-$linear equations. However, we simplify the problem and follow the leading order terms in the next section.
\section{Linearized Einstein$-$Gauss$-$Bonnet's equations }
In this section we simplify the complicated setup of non$-$linear Einstein$-$Gauss$-$Bonnet's equations by linearizing them around the final AdS Gauss$-$Bonnet black brane solution \eqref{GBmetric}. As it was explained, the linearizing of these equations is interpreted as an amplitude expansion on top of the black brane. By considering the parameter of the expansion as $\alpha$, one expands the metric functions similar to \eqref{linearizedmetricfuncs} as follows

\begin{subequations}
\label{linearizedmetricfuncs}
\begin{eqnarray}
\label{linearizedGB.a}
A\left(t,z\right)&=& \frac{1}{4 \lambda _{GB} z^2}\left(1-\sqrt{1-8\lambda _{GB}(1-\frac{z^4}{z_h^4})}\,\,\right) +\alpha \delta A^{(1)}(t,z)+\mathcal{O}\left(\alpha ^2 \right) \,, \\ [2mm]
\label{linearizedGB.b}
B\left(t,z\right)&=&\alpha \delta B^{(1)}(t,z)+\mathcal{O}\left(\alpha ^2 \right)  \,, \\ [2mm]
\label{linearizedGB.c}
\Sigma\left(t,z\right) &=& \frac{1}{z}+\alpha \delta \Sigma ^{(1)}(t,z)+\mathcal{O}\left(\alpha ^2 \right) \,. \,\,\,\,\,\,\,
\end{eqnarray}
\end{subequations}
Inserting these relations to Einstein-Gauss-Bonnet's equations, we find that $\delta A^{(1)} \left(t,z\right)$ and $\delta \Sigma ^{(1)} \left(t,z\right)$ vanish and we have an evolution equation for $\delta B ^{(1)}\left(t,z\right)$ as

\begin{align}\label{32}
&\left(z^4+3\right) \partial_z\delta B+z \left(z^4-1\right) \partial^2_z\delta B-3 \partial_t\delta B+2 z \partial_t\partial_z\delta B+\lambda_{GB} \big\{\left(6 z^9-2
z\right) \partial^2_z\delta B\nonumber\\&+6 \left(5 z^8+1\right) \partial_z\delta B+4 \left(z^4-3\right) \partial_t\delta B+8 z \left(z^4+1\right) \partial_t\partial_z\delta B\big\}=0.
\end{align}

\vspace{0.3cm}

\begin{figure}[ht]
\includegraphics[width=5.2cm]{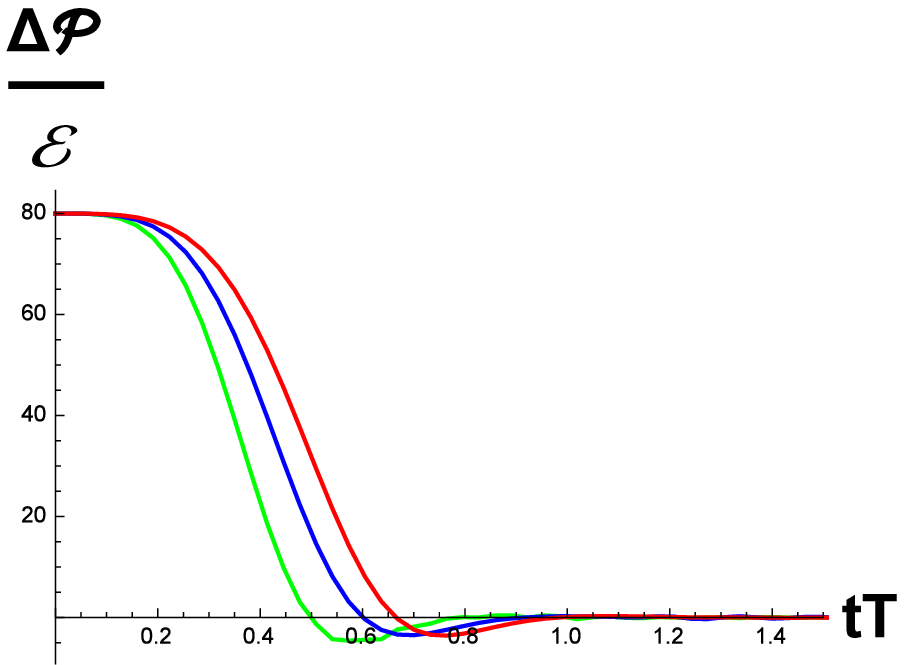} \includegraphics[width=5.2cm]{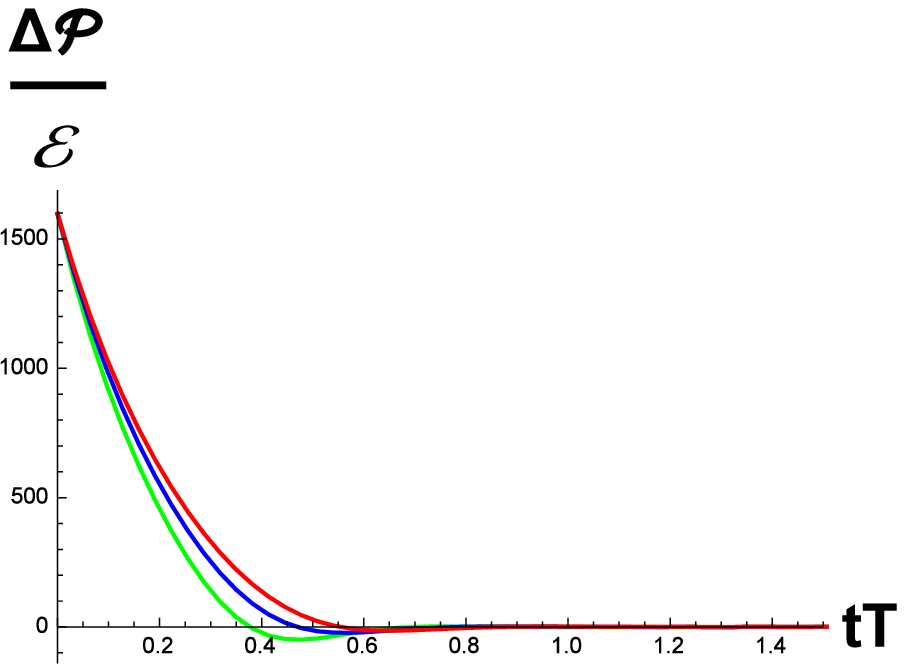}\includegraphics[width=5.2cm]{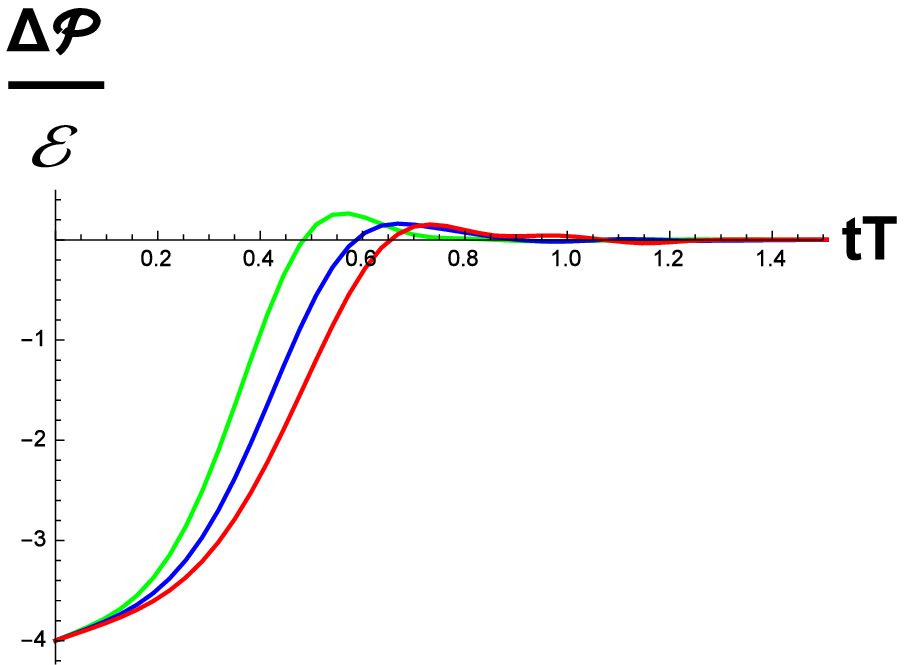} \\ \hfill
\\ \includegraphics[width=5.2cm]{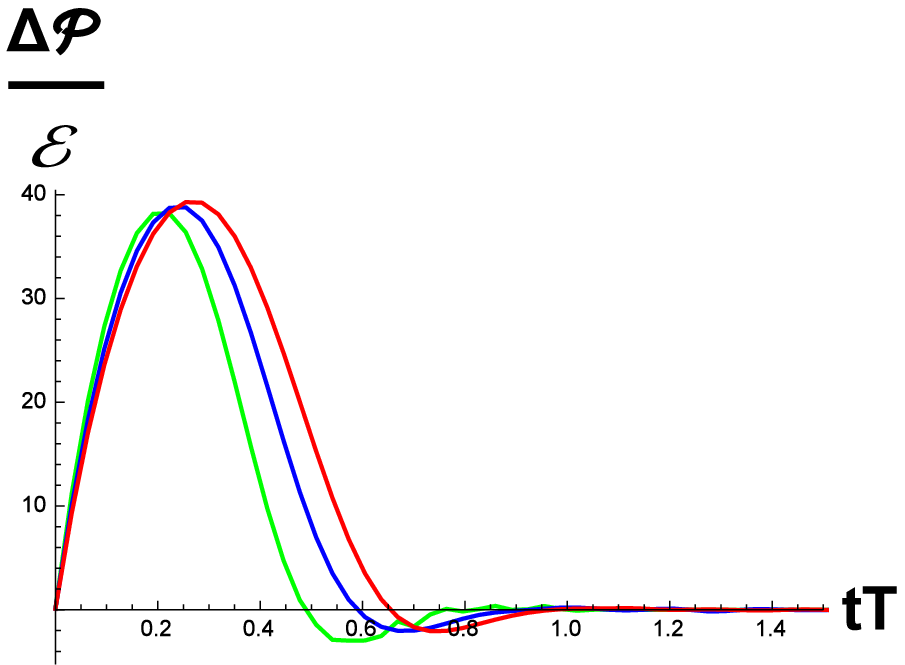} \includegraphics[width=5.2cm]{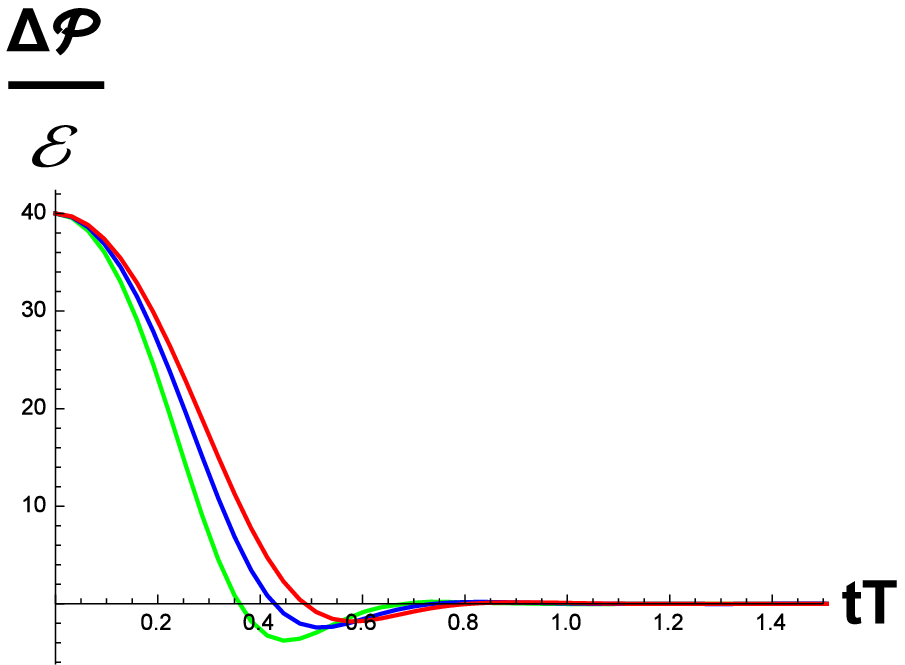}\includegraphics[width=5.2cm]{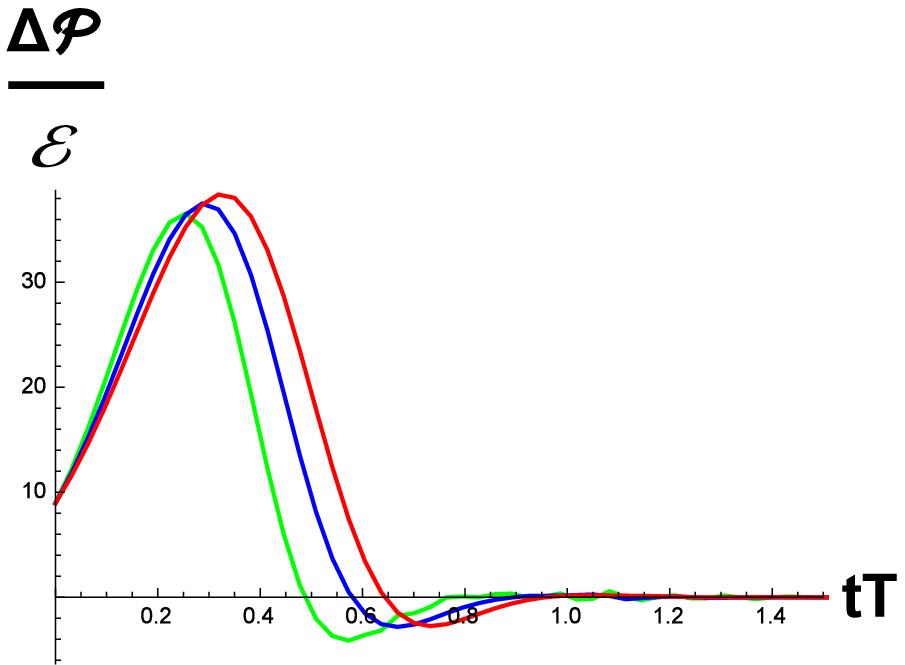} \\ \hfill
\\ \includegraphics[width=5.2cm]{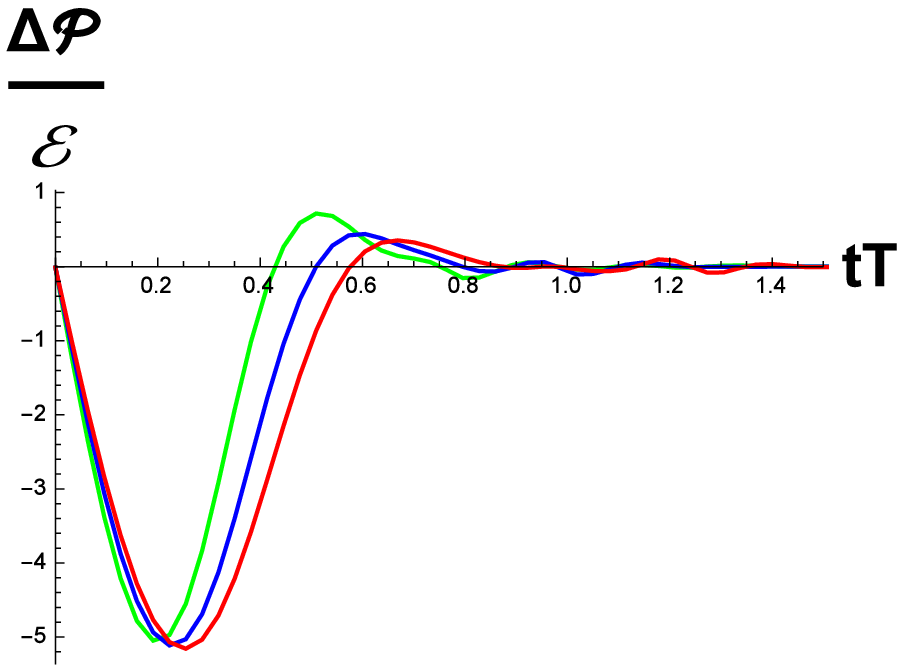} \includegraphics[width=5.2cm]{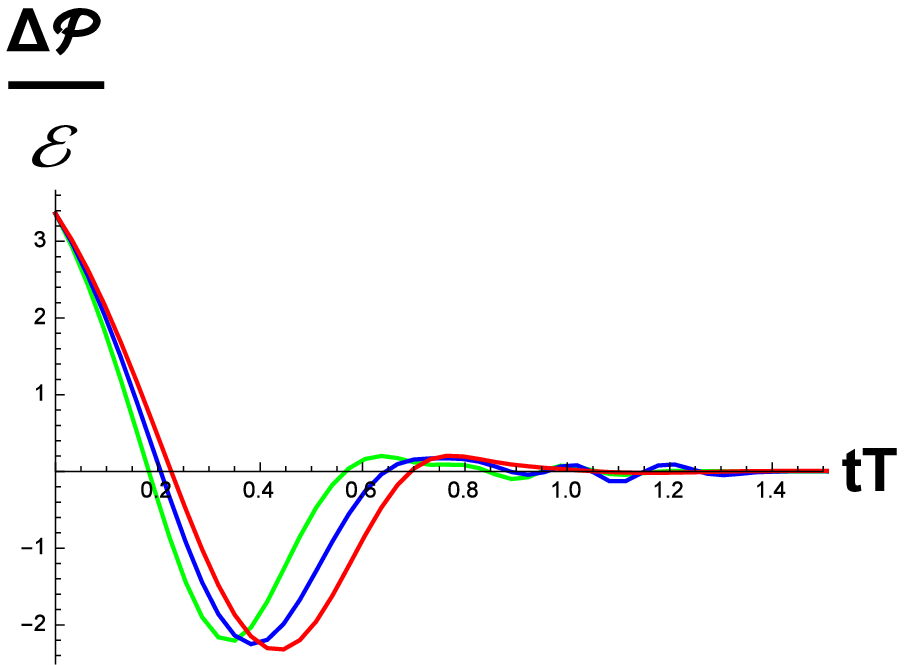}\includegraphics[width=5.2cm]{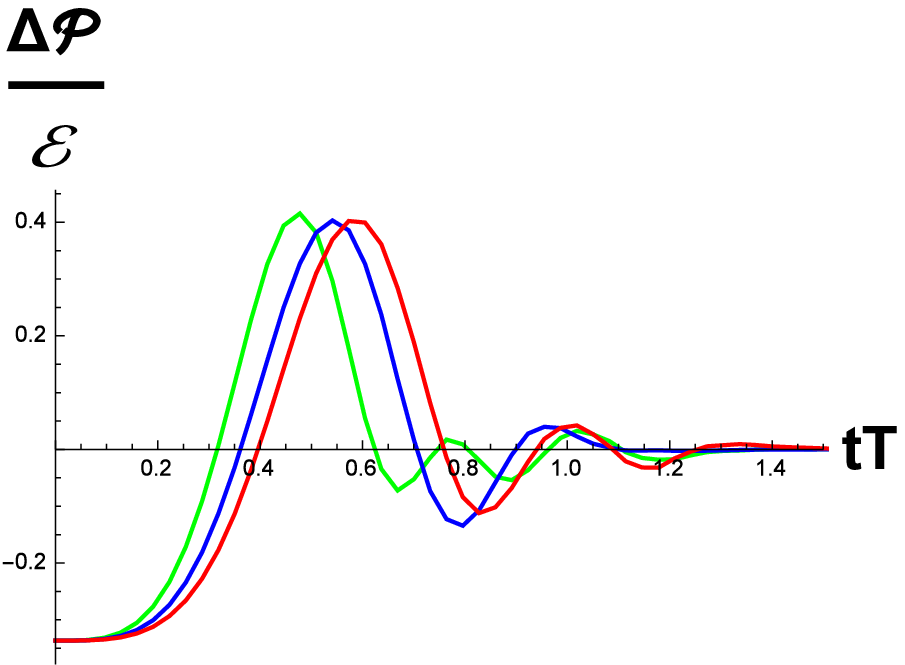} \\ \hfill
\caption{\label{f2} The ratio of  $\frac{\Delta \mathcal{P}}{\mathcal{E}}$ vs $tT$ for nine different initial conditions at first order linearization. The Red, Blue and Green curves correspond to  $\lambda _{GB}=-0.05,0.0,0.05$ respectively.}
\end{figure}

In the case of $\lambda _{GB}=0$, it changes to the equation \eqref{Bequation}.\\

To solve the above equation, one should regularize $B(t,z)$ as \eqref{regforB} and the solution must satisfy the condition $\delta B_{reg}^{(1)}\left(t,z=0\right)=0$. The energy density $\mathcal{E}$ is constant in this setup. However, a precise field theory dual to the Gauss$-$Bonnet gravity is unknown and we only study the ratio $\Delta \mathcal{P}(t)/\mathcal{E} $. \\

Now we give a detailed discussion of the comparison of initial data with different Gauss-Bonnet coupling constant. First, one finds from  equation (\ref{24}) that the following quantity depends on the $\lambda_{GB}$,
\be
\Delta \mathcal{P}(t)/\mathcal{E} =\frac{-2b_4(t)}{L_C^2},
\ee
where we fixed the energy density by $a_4=-1$. Next, according to the near boundary expansion (\ref{nearbdryexpansionsgb}) and linear approximation in (\ref{linearizedmetricfuncs}) one finds the ratio as
\be
\Delta \mathcal{P}(t)/\mathcal{E} =-2 \frac{\alpha \delta B}{z^4} (1+\lambda _{GB}).
\ee

To have a meaningful comparison of different initial states with different $\lambda _{GB}$, we forced that the $\Delta \mathcal{P}(t)/\mathcal{E} $ quantity is $\lambda _{GB}-$independent at initial time. In this way, by changing $\lambda _{GB}$ the initial states starts from the same value. Technically, we apply the following condition at initial time $t=t_{ini}$ as
\be
\label{BBc}
\delta B_{\lambda _{GB}}\left(z,t=t_{ini}\right)=\frac{1}{1+\lambda _{GB}} \delta B_{\lambda _{GB}=0}\left(z,t=t_{ini}\right)\,.
\ee

In Fig. \ref{f2}, we plot the ratio of  $\frac{\Delta \mathcal{P}}{\mathcal{E}}$ as a function of $tT$ where $t$ is time process and $T$ is the final equilibrium temperature. The Red, Blue and Green curves correspond to  $\lambda _{GB}=-0.05,0.0,0.05$, respectively. We have analyzed a large number of initial states to understand the effect of $\lGB$. The nine different initial non$-$equilibrium states have been shown in this figure. As it is clear, all states initiated at a common point. \\

We find that considering $\lGB$ does not change the early times behavior of the pressure anisotropy. This observation expected from the fact that the near boundary dynamics of  $\delta B(t,z)$ is approximately linear. It only differs at transient time because in this case the signal propagates from the interior of the bulk gravity not from the boundary. Also, one finds that the general features of the plots do not change by considering different values of Gauss-Bonnet coupling $\lGB$. One concludes from Fig. \ref{f2} that there is a shift for isotropization plots by considering different signs of $\lGB$. This could be a key result of our study. \\

\begin{figure}[ht]
\includegraphics[width=7.5cm]{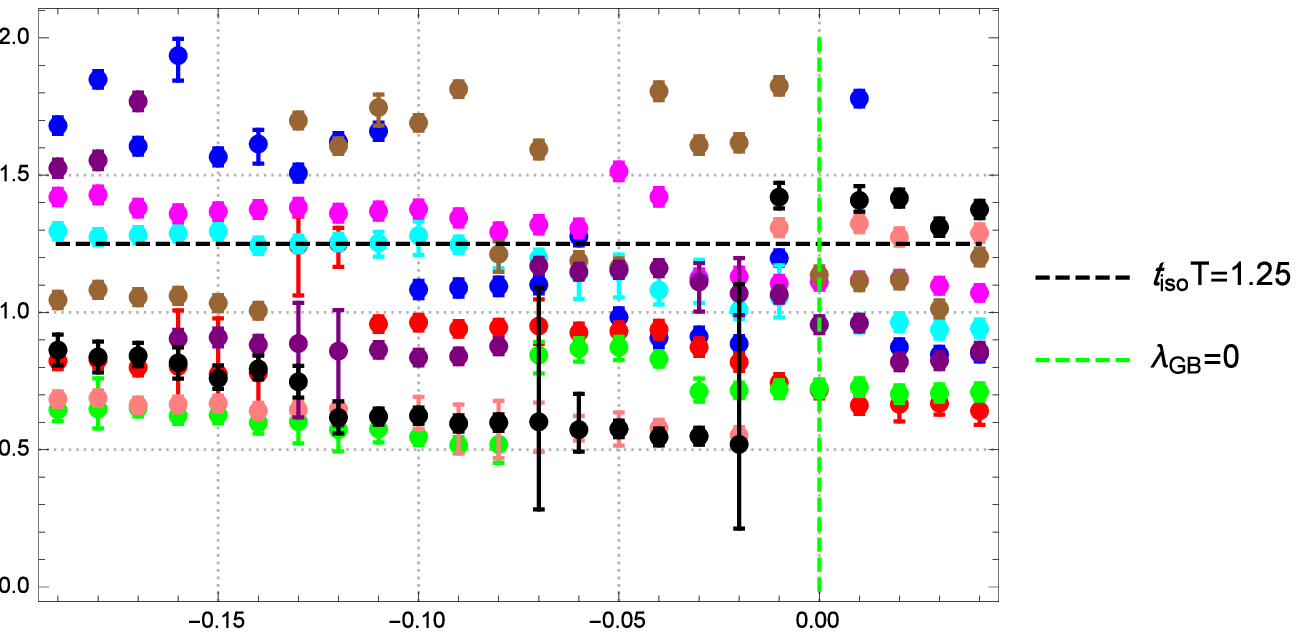}\includegraphics[width=7.5cm]{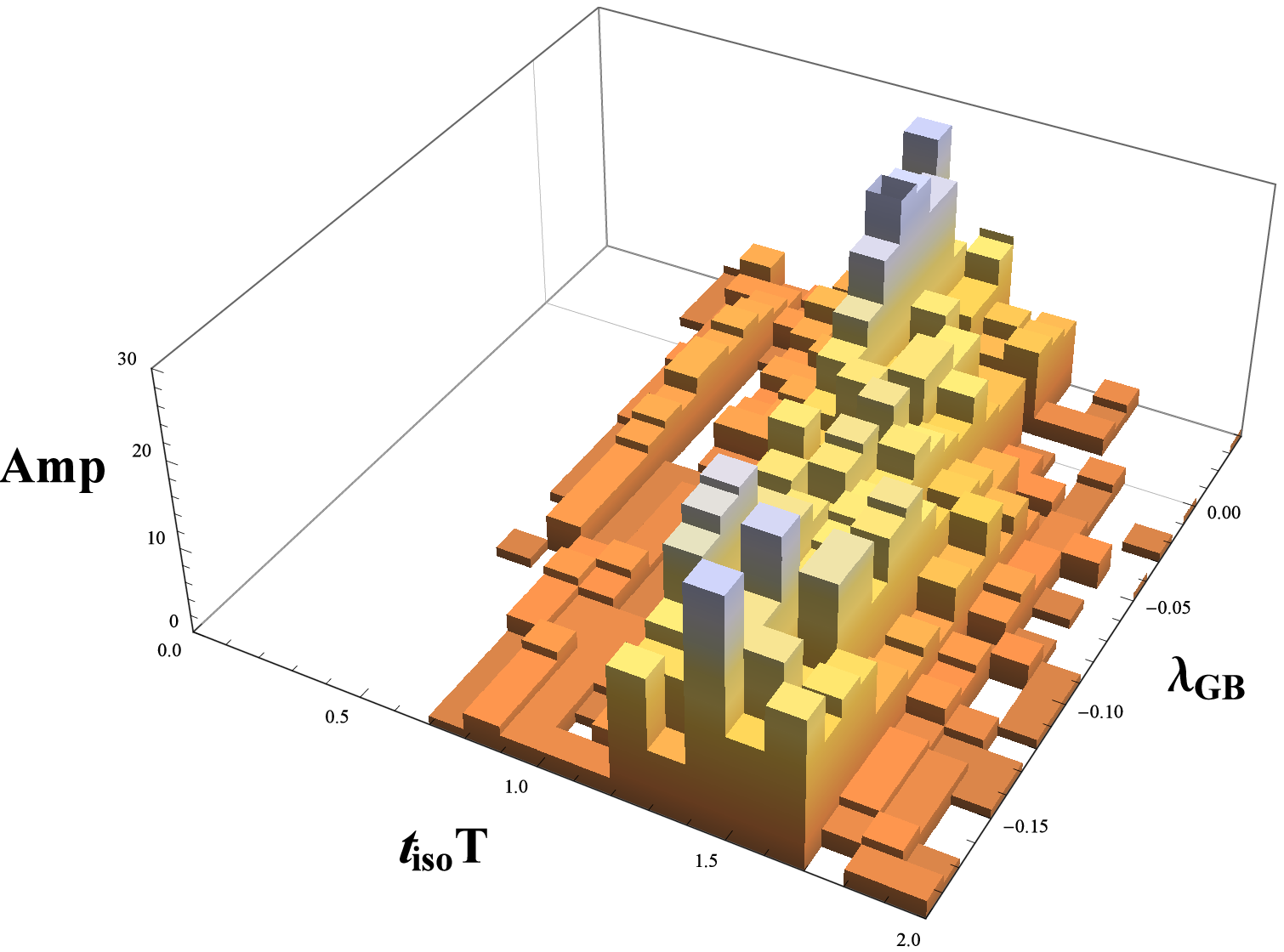}
\caption{\label{f5} Left plot: The $t_{iso}T$ vs $\lambda _{GB}$ for nine initial conditions. Different colors correspond to different initial conditions. We set isotropization time as time that the ratio $\frac{\Delta \mathcal{P}}{\mathcal{E}}$ becomes smaller than $0.1$. The error bars are difference between selected isotropization time and time that  $\frac{\Delta \mathcal{P}}{\mathcal{E}}<0.1 \pm 0.02$.
Right plot: Histogram for $t_{iso}T$ vs $\lambda _{GB}$ for about 100 initial non$-$equilibrium states.
 For  $\lambda _{GB}=0$, the $t_{iso}T$ quantity is smaller than   $1.25$ which means $\mathcal{O}(1)$ for all of initial states (That it is agreed with results of \cite{Heller:2013oxa}). For  $\lambda _{GB}\neq 0$, there are some initial conditions that corresponding $t_{iso}T$ is  $\mathcal{O}(>1)$.}
\end{figure}

In Fig. \ref{f5}, we increase the number of the non$-$equilibrium states and explicitly show that how the $t_{iso}$ changes by $\lGB$. In the left plot of this figure, different colors correspond to different initial states.  We set isotropization time as time that the ratio $\frac{\Delta \mathcal{P}}{\mathcal{E}}$ becomes smaller than $0.1$. The error bars show difference between selected isotropization time and time that $\frac{\Delta \mathcal{P}}{\mathcal{E}}<0.1 \pm 0.02$.

In the right plot of this figure, a histogram plotted for $t_{iso}T$ as a function of $\lambda _{GB}$. One finds that $t_{iso}T$ is smaller than $1.25$ for $\lambda _{GB}=0$  which means $\mathcal{O}(1)$ for all of initial states (That it is agreed with results of \cite{Heller:2013oxa}). Interestingly, for $\lambda _{GB}\neq 0$, there are some initial states that corresponding $t_{iso}T$ is greater than $1.25$.

\section{The entropy production}
In this section we investigate effect of finite coupling corrections on the entropy production during the isotropization process. The motivation is to study a quantity which depends on the IR geometry.

As it was argued in \cite{Heller:2013oxa}, the entropy production can be studied by considering quadratic corrections to linearized Einstein's equations. Then study of time evolution of $\delta\,A^{(2)}$ and $\delta\,\Sigma^{(2)}$ becomes important. One should notice that at linear order the entropy production does not change and one should extend the linear analysis to second order corrections.

Regarding \cite{Heller:2013oxa}, we define the entropy production of initial non$-$equilibrium states from the event horizon. Although the definition is only relevant to the near equilibrium not far from equilibrium situation. Notice that there is no guarantee for increasing the entropy, see \cite{Heller:2013oxa} for more details. Ref. \cite{Heller:2013oxa}, finds very good agreement with non$-$linear result and show that a $20$ percent accuracy.\\

The event horizon is defined as follows
\begin{equation}
    r-r_{eh}(t)=0,
\end{equation}
where
\begin{equation}\label{34}
    r'_{eh}(t)-\frac{1}{2}A\left(t,r_{eh}(t)\right)=0,
\end{equation}

In the asymptotic future, the geometry goes to AdS Gauss-Bonnet black brane in \eqref{GBmetric} and $r_{eh}(\infty) \to \pi T$ where $T$ is the equilibrium temperature at finite coupling in \eqref{TGB}. The entropy is proportional to the area of the event horizon as
\begin{equation}\label{35}
    S_{eh}(t)\propto \Sigma\left(t,r_{eh}(t)\right)^3,
\end{equation}
To find $\delta\,\Sigma^{(2)}$, one can perturb \eqref{em1} to second order as
\begin{align}\label{36}
&-({\partial_z\delta
	B^{(1)}})^2-4 {\partial_z\delta \Sigma^{(2)}}-2\, z\, {\partial_z^2\delta \Sigma^{(2)}}+{\lambda_{GB}} \big\{4 \,z\, {\partial_z^2\delta
	B^{(1)}} {\partial_t\delta
	B^{(1)}}\nonumber\\&+4 \left(z^4-1\right)
\left(({\partial_z\delta B^{(1)}})^2+z
{\partial_z^2\delta B^{(1)}} {\partial_z\delta
	B^{(1)}}+4 {\partial_z\delta \Sigma^{(2)}}+2 \, {\partial_z^2\delta \Sigma^{(2)}}\right)\big\}=0
\end{align}
At $\lGB=0$, one finds the same equation in \cite{Heller:2013oxa}. This is an ordinary differential equation and can be sloven on each time slice to find $ \delta B^{(1)} $. Therefore, we can use the above equation to determine $ \delta \Sigma^{(2)} $ from $\delta B^{(1)} $.

By expansion of \eqref{em5} to second order, we find the following differential equation, which for simplicity we have written it to linear order in $ \lGB $ expansion as
\begin{align}\label{37}
&-3 {\delta A^{(2)}}-\frac{3 \partial_t{\delta B^{(1)}} \partial_z{\delta B^{(1)}}}{2}-6 \partial_t{\delta \Sigma^{(2)}}+\partial_z{\delta A^{(2)}} z-\frac{3}{4} (\partial_z{\delta B^{(1)}})^2 \left(z^4-1\right)\nonumber\\& -6 \partial_z{\delta \Sigma^{(2)}} \left(z^4-1\right)+\frac{\partial_z^2{\delta A^{(2)}} z^2}{2}-\frac{6 {\delta \Sigma^{(2)}} \left(z^4-1\right)}{z}+\lambda _{{GB}} \bigg\{6 (\partial_t{\delta B^{(1)}})^2\nonumber\\&+\partial_z^2{\delta B^{(1)}} \left(2 \partial_t{\delta B^{(1)}} \left(1-3 z^4\right) z+2 \partial_t^2{\delta B^{(1)}} z^2\right)-24 \partial_t{\delta \Sigma^{(2)}} \left(5 z^4+1\right)\nonumber\\&-2 z^2(\partial_t\partial_z{\delta B^{(1)}})^2 +\partial_z{\delta B^{(1)}} \big(-2 \partial_t{\delta B^{(1)}} \left(13 z^4+9\right)+4 \partial_t\partial_z{\delta B^{(1)}} z \left(1-3 z^4\right)\nonumber\\&-2 \partial_z^2{\delta B^{(1)}} z \left(3 z^8-4 z^4+1\right)\big)-4 \partial_t{\delta B^{(1)}} \partial_t\partial_z{\delta B^{(1)}} z-4 \partial_z{\delta A^{(2)}} z \left(z^4-1\right)\nonumber\\&+\frac{1}{2} (\partial_z{\delta B^{(1)}})^2 \left(-53 z^8+16 z^4+15\right)+12 \partial_z{\delta \Sigma^{(2)}} \left(-11 z^8+8 z^4+1\right)\nonumber\\&-2 \partial_z^2{\delta A^{(2)}} z^2 \left(z^4-1\right)-12 {\delta A^{(2)}} \left(5 z^4+1\right)+\frac{12 {\delta \Sigma^{(2)}} \left(-11 z^8+8 z^4+1\right)}{z}\bigg\}=0.
\end{align}
By sending $\lGB$ to zero, one finds the related equation in  \cite{Heller:2013oxa}. Notice that by finding $ \delta \Sigma^{(2)} $ and
$ \delta B^{(1)} $, one finds $ \delta A^{(2)} $ from the above equation. To improve the accuracy of the numerics, they should be redefined as
\begin{equation}
 \delta A_{reg}^{(2)}=\frac{ \delta A^{(2)}}{z^4},\,\,\,\,\ \delta \Sigma_{reg}^{(2)} = \frac{ \delta \Sigma^{(2)}}{z^5}.
\end{equation}
\begin{figure}[ht]
\includegraphics[width=15cm]{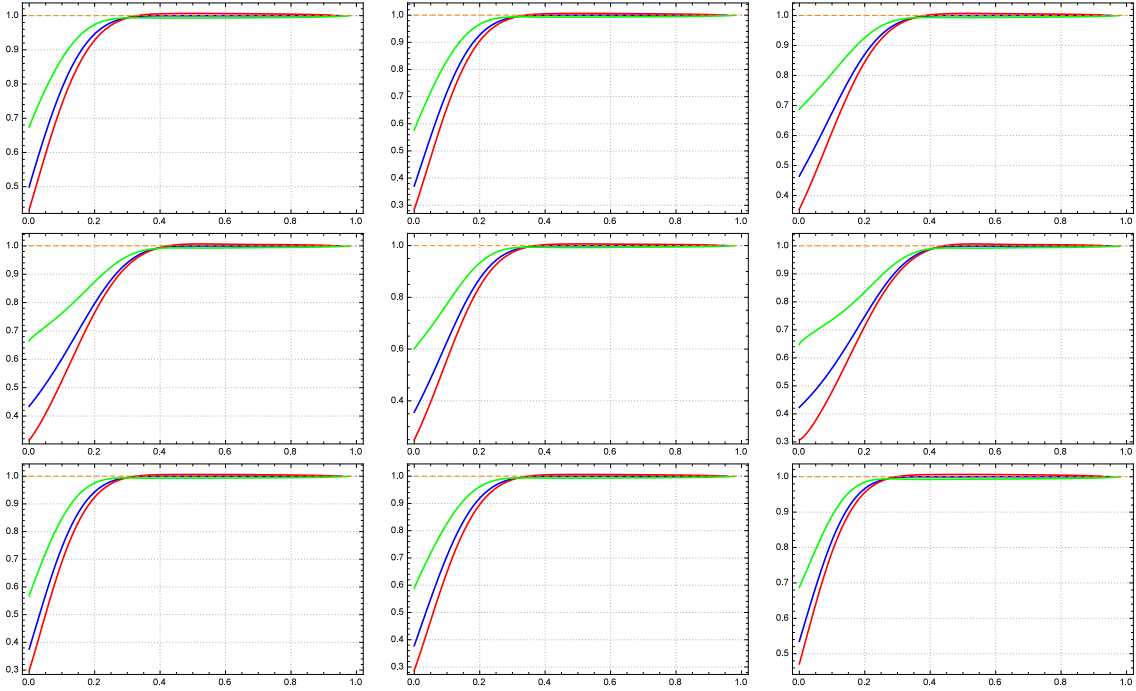}
\caption{\label{f3} The ratio of  $\frac{ \mathcal{S}_{eh}(t)}{\mathcal{S}_{eh}(\infty)}$ vs $tT$ for nine different initial non$-$equilibrium states at second order linearization. The Red, Blue and Green curves correspond to  $\lambda _{GB}=-0.05,0,0.05$, respectively. }
\end{figure}

The numerical techniques are much simpler than the non$-$linear case. Briefly, one can itemize the computing event horizon entropy as below:
\begin{enumerate}
\item Solve the  \eqref{32} for $\partial _t \delta B^{(1)}(t,z)$ as evolution equation of  $\delta B^{(1)}(t,z)$.
\item Insert the initial state of  $\delta B^{(1)}(t,z)$ into \eqref{36}, and find  $\delta \Sigma^{(2)}(t,z)$ for each time slice.
\item Having both  $\delta B^{(1)}(t,z)$ and  $\delta \Sigma^{(2)}(t,z)$, and solving \eqref{37}, one can find  $\delta A^{(2)}(t,z)$ in each time step \footnote{We used spectral method to perform steps 1 to 3. }.
\item Repeat the three above steps for subsequence time slices.
\item Solve the first order ODE \eqref{34} to determine evolution of event horizon \footnote{Using  \emph{4-th order Runge-Kutta(RK4)} algorithm and \emph{"Interpolation"} command of \emph{Mathematica} is sufficient to solve that ODE.}
\item Compute the event horizon entropy by \eqref{35}.
\end{enumerate}

Again, to have a meaningful comparison of different initial states with different $\lambda _{GB}$, we consider the $\Delta \mathcal{P}(t)/\mathcal{E} $ quantity to be $\lambda _{GB}-$independent at initial time. Numerically, we used the condition \eqref{BBc}.

Results for entropy production is shown in Fig. \ref{f3}. In this figure, the ratio of  $\frac{ \mathcal{S}_{eh}(t)}{\mathcal{S}_{eh}(\infty)}$ as a function of $tT$ for nine different initial states has been studied. The Red, Blue and Green curves correspond to  $\lambda _{GB}=-0.05,0,0.05$, respectively.\\

One finds that Gauss$-$Bonnet coupling $\lGB$ has an important effect at early time process of isotropization. Depending on the sign of $\lGB$, the entropy production changes. For $\lGB\,>\,0$, the entropy production increases while for $\lGB\,<\,0$ it decreases. At transient times, one finds an especial time where the entropy does not dependent on the values of $\lGB$. It is interesting that this time is less than the isotropization time. It is questionable if such behavior exist also by studying non$-$linear equations.

\section{Discussion }
In this paper, we studied holographic isotropization of an anisotropic homogeneous non$-$Abelian strongly coupled in the presence of Gauss$-$Bonnet coupling corrections. As a general result of the AdS/CFT correspondence, the effects of
finite but large 't Hoof coupling in the boundary gauge field theory is related
to higher derivative terms in the corresponding geometry.

In this paper, we considered Gauss-Bonnet higher derivative terms. Such
curvature squared terms are common from the sense that comes from string theory and do not have usual difficulties with considering higher derivative terms. The Gauss$-$Bonnet theory is an example
of more general Lovelock theories where the usual difficulties of considering higher
derivative terms like instability is absent. Therefore they are interesting for studying
non$-$perturbative effects in the presence of higher derivative corrections.\\

For the first time, we derived the nested Einstein$-$Gauss$-$Bonnet's equations in \eqref{GBa1},\eqref{GBb1},\eqref{em5},\eqref{GBd1} and \eqref{em1}. It was verified that one can linearize Einstein's equations around the final black
hole background. Using this observation, we simplified the complicated setup and studied the expectation value of the boundary stress tensor. An understanding of how the isotropization process of a non$-$Abelian strongly coupled plasma is affected by considering finite coupling corrections may be essential for theoretical predictions. The main motivation for our study is to see if the fast thermalization
depends on these corrections. \\

One of the main results of this paper is that the thermalization time increases at finite coupling. We studied the
isotropization times of some non$-$equilibrium states at finite coupling in Fig. 2. We find that considering $\lGB$ does not change the early times behavior of the pressure anisotropy. As a key result of our study, it is shown that there is a shift for isotropization plots by
considering different signs of $\lGB$, see Fig. 1. \\

We also studied the entropy production in the presence of Gauss$-$Bonnet corrections. This is a quantity which depends on the IR bulk
geometry. To study this observable, we considered quadratic corrections to linearized
Einstein$-$Gauss$-$Bonnet's equations. It is found that at early times of the isotropization process the entropy production increases for $\lGB>0$ and decreases for $\lGB<0$. It is found that at transient times, which is smaller than the isotropization time, there is an especial time where the entropy does not depend on the different values of $\lGB$.

It is an important question if the above results also exist in the case of non$-$linear Einstein$-$Gauss$-$Bonnet's equations. We leave this interesting problem as a future work.

\section*{Acknowledgments}
This project started in the workshop "Lecture series on AdS Numerics, 8-14 June 2015, IPM Tehran, Iran". M. A and K. B. F would like to thank the organizers M. Ali-Akbari and H. Ebrahim, and especially very grateful to Wilke Van der Schee for giving these lectures and more explanations about writing the numerical codes. We also thank Michal P. Heller for useful comments on the initial states and boundary conditions. Also thank M. H. Vahidinia, F. Charmchi and L. Shahkarami for useful discussions. K.B.F also thanks the organizers of "Conference on Non-Equilibrium Phenomena
in Condensed Matter and String Theory, 30 June - 4 July 2014, ICTP" and ICTP for hospitality. Symbolic tensor calculations has been carried out by Mathematica package xAct \cite{Nutma:2013zea}.


\end{document}